\documentclass[a4paper,12pt]{article}

\usepackage{amsmath,amssymb,graphicx}
\usepackage{cite}

\numberwithin{equation}{section}

\newcommand{\ii}{\mathrm{i}}

\newcommand{\dd}{\mathrm{d}}

\newcommand{\tr}{\mathop{\mathrm{tr}}\nolimits}

\newcommand{\A}{\mathcal{A}}

\newcommand{\I}{\mathbb{I}}

\newcommand{\ft}[2]{{\textstyle\frac{#1}{#2}}}
\newcommand{\qp}[1]{[\!\![ #1 ]\!\! ]}

\newtheorem{lemma}{Lemma}
\begin{document}

\title{On Nambu-Lie 3-algebra representations}%
\author{
Corneliu Sochichiu\thanks{On leave from~: Institutul de
    Fizic\u a Aplicat\u a A\c S, str. Academiei, nr. 5,
    Chi\c{s}in\u{a}u, MD2028 Moldova; e-mail:
\texttt{sochichi@sogang.ac.kr}}\\
{\it Center for Quantum Spacetime, }\\
{\it Sogang University,}\\
{\it  Seoul 121-742, Korea}
}%

%

\maketitle
\begin{abstract}
 We propose a recipe to construct matrix representations of Nambu--Lie 3-algebras in terms of irreducible representations of underlying Lie algebra. The case of Euclidean four-dimensional 3-algebra is considered in details. We find that representations of this 3-algebra are not possible in terms of only Hermitian matrices in spite of its Euclidean nature.
\end{abstract}
\maketitle
\section{Introduction}

Recent proposals for the theory describing multiple M2 branes \cite{Bagger:2006sk,Bagger:2007jr,Bagger:2007vi,Gustavsson:2007} revived the interest in structure appearing in Nambu mechanics \cite{Nambu:1973qe}. A particularly dramatic issue is related to Nambu--Lie algebras \cite{Gustavsson:2007,Bandos:2008,Ho:2008a,Papadopoulos:2008c,%
Gauntlett:2008,Papadopoulos:2008a,Ho:2008ve,Krishnan:2008zm,%
Passerini:2008,Cecotti:2008qs}. Already in \cite{Nambu:1973qe} it was noted that there are only very restricted possibilities for certain types of a non-trivial Nambu--Lie 3-algebra.  This subject was further developed by  \cite{Takhtajan:1993vr,Chatterjee:1996,Dito:1997,Chatterjee:1996a}.
Thus the only ``compact'', i.e. having positive scalar product finite dimensional algebra was found to be the four-dimensional algebra $\A_4$ with structure constants given by the totally anti-symmetric four-dimensional tensor\cite{Papadopoulos:2008,Gauntlett:2008}.\footnote{There also exist less restrictive higher order algebras called Generalized Lie Algebras, based on the Schouten-Nijenhuis bracket \cite{deAzcarraga:1996jk} classified in \cite{DeAzcarraga:1996ts}. For their comparison to Nambu--Lie 3-algebras considered here see \cite{deAzcarraga:1997uc}.}

On the other hand, for the viability of the multiple M2-brane model we need an infinite family of algebras parameterized by an integer parameter, which can be interpreted as the number of M2-branes.

In fact, giving up the Euclidean metric or finite dimension requirement allows one to construct various 3-algebras from Lie algebras by an uplift procedure. It is clear that all such 3-algebras are either Lorenzian or infinite-dimensional \cite{FigueroaO'Farrill:2008zm,Gomis:2008a,Lin:2008qp}.

In this work we propose a construction of representations of Nambu--Lie 3-algebras in terms of matrix operators subject to Nambu commutation relations. The construction is based on the fact that the Nambu--Lie 3-algebra includes a set of underlying Lie algebra structures, which are obtained by fixing a Nambu--Lie 3-algebra element. Then the Nambu--Lie algebra representation should be split into irreducible representations of the underlying Lie algebra. Choosing such Lie algebra representations from which the 3-algebra representation (re)constructed solves the problem.

The case of four-dimensional Euclidean 3-algebra $\A_4$ is considered here in more details. It appears that this 3-algebra can not be represented, in spite of its Euclidean nature, in terms of sole Hermitian matrices. The uplift of a representation of a compact Lie algebra requires complex eigenvalues for the fixed element parameterizing the underlying Lie algebra. Although it is not clear wether the setup of our construction is the most general one, it seems that this feature of 3-algebra representations is quite general.

In spite of the fact that the role of representations of 3-algebras in the theory of multiple M2-branes is not yet clear, we hope that understanding the representations may be useful also for a better understanding of the latter.

The plan of the paper is as follows. In the next section we define the matrix representation for a 3-algebra and describe the requirements it should satisfy. Next, we give a general construction of such a representation. After that we specify to algebra $\A_4$ for which we find the explicit form of representations. Then we analyze the two-dimensional representation in most general context, but find that this case too in fact reduces to the previous one.

\section{Lie algebra induced representation}

\subsection{Objectives}
Consider a Nambu--Lie 3-algebra defined as the linear span of generators $\{T_a\}$, $a=1,\dots, D$ subject to the following 3-bracket relation,
\begin{equation}\label{nambuSU2}
  [T_a,T_b,T_c]=\ii f_{abc}{}^{d}T_d,
\end{equation}
where $f_{abc}{}^{d}$ are the structure constants, which are anti-symmetric in the lower indices and subject to Fundamental Identity,
\begin{equation}
  f_{a_1 a_2 ba}f_{a_3 a_4 a_5 b}=
  f_{a_1 a_2 a_3 b}f_{b a_4 a_5a}+
  f_{a_1 a_2 a_4 b}f_{a_3 b_2 a_5a}+
  f_{a_1 a_2 a_5 b}f_{a_3 b_2 a_5a},
\end{equation}
for all $a$. The indices are lowered and raised respectively by the Killing metric $h_{ab}$ and its inverse. Next we chose the basis in which $h_{ab}=\delta_{ab}$. In this basis the structure constants are totally anti-symmetric. Also we suppose that the structure constants $f_{abc}{}^{d}$ are non-degenerate in the sense that there exist ``dual structure constants'' $\tilde{f}^{abcf}$ such that,
\begin{equation}
  \tilde{f}^{abc}{}_f f_{abc}{}^{d}=\delta_f^d.
\end{equation}
This is not a very constraining requirement, in any case any semi-simple Lie algebra satisfies an analogue of this. In the case of totally anti-symmetric structure constants $f$, the dual constants can be chosen to be anti-symmetric too.

Our aim is to construct a matrix representation $R(\A)$ such that the Nambu-Lie bracket is mapped to the Nambu commutator \cite{Nambu:1973qe},
\begin{equation}\label{NambuC}
  [A,B,C]_{N}=ABC+BCA+CAB-BAC-ACB-CBA,
\end{equation}
where $A,B,C\in R(\A)$.

\subsection{General construction}
As is known a Nambu-Lie $n$-algebra generates a family of lower degree Nambu-Lie algebras. In the case of 3-algebra one has a family of Lie brackets parameterized by an element $\vec{\xi} \in\A$,
\begin{equation}\label{eq:NL2L}
  [g,f]_{\vec{\xi}}\equiv[\vec{\xi},g,f],\qquad \vec{\xi},g,f \in \A.
\end{equation}
As pointed in \cite{Takhtajan:1993vr}, the restricted bracket \eqref{eq:NL2L} is anti-symmetric in $g$ and $f$ and satisfies the Bianchi identity, therefore a Nambu--Lie algebra contains incorporated Lie algebra structures.

In the case when the 3-algebra basis is not fixed or it is invariant with respect to rotations (as in the case of the four-dimensional algebra considered below), we can chose $\xi$ to be along $D$-direction: $\vec{\xi}=\xi T_{D}$. Since $\xi$ enters homogeneously in \eqref{NambuC}, its value is not relevant, therefore we can generically set $\xi=1$. Then the underlying $(\dd-1)$-dimensional Lie algebra is defined by the commutation relations,
\begin{equation}\label{eq:NLreduced}
  [T_i,T_j]_{t_D}\equiv [T_D,T_i,T_j]=\ii f_{ijk}T_k,
\end{equation}
where $f_{ijk}=f_{D ijk}$, $i,j,k=1,\dots, D-1$.

Since we seek a ``unitary'' representation of 3-algebra, it is natural to assume, that the the matrix representing $T_D$ can be diagonalized within this representation.
Consider the subspace corresponding to the eigenvalue $t_D$. The Nambu--Lie commutator on such a space reduces to\footnote{Note that here we abusively denote both algebra generators and the their representation as $T_i$.}
\begin{equation}\label{eq:tDLie}
  [T_i,T_j]_{t_D}=t_D[T_i,T_j],
\end{equation}
where $[\cdot,\cdot]$ is the usual matrix commutator. In order to satisfy the Nambu--Lie commutation relations \eqref{eq:NLreduced}, the reduced representation on $t_D$ subspace should satisfy the following matrix commutation relations,
\begin{equation}
  [T_i,T_j]=\ii t_D^{-1}f_{ijk}T_k.
\end{equation}

Let us consider the situation when the underlying algebra is a semi-simple Lie algebra. Then the $t_D$-eigen-space is split into irreducible representations of the semi-simple algebra. The irreducible representations of the algebra \eqref{eq:tDLie} correspond to eigen-spaces of the quadratic Casimir operator,
\begin{equation}
  C_2(T)=t_D^{-2}c_2(\tau),
\end{equation}
where $c_2(\tau)$ is the quadratic Casimir of the $t_D$-independent algebra with standardized commutation relations,
\begin{equation}\label{eq:su2standard}
  [\tau_i,\tau_j]=\ii f_{ijk}\tau_k,
\end{equation}
where $\tau_i=t_D T_i$.

Now let us pick up a particular irreducible representation and let us compute the following quantity restricted to this representation,
\begin{equation}\label{consistency}
 t_D \equiv T_D=-\ii\, \tilde{f}^{ijk}{}_D[T_i,T_j,T_k]_N
 =3t_D^{-3} \tilde{f}^{ijk}{}_Df_{jkl}\tau_i\tau_l
  ,
\end{equation}
where we make use of the total anti-symmetric property of $\tilde{f}^{abc}{}_d$. This implies that the value $t_D$ of the matrix $T_D$ should satisfy,
\begin{equation}\label{eq:tD4}
  t_D^4=\tilde{f}^{ijk}{}_Df_{jkl}\tau_i\tau_l,
\end{equation}
where the r.h.s. depends only on the details of the representation of the Lie algebra and should be related to the value of Casimir operator.
So, the eigenvalue of $T_4$ is expressed as the fourth power root from the r.h.s of \eqref{eq:tD4}. It is clear, that if the r.h.s fails to be positive, the representation can not be realized in terms of Hermitian matrices.

In the case in which the r.h.s. of \eqref{eq:tD4} contains beyond the Casimir also a Lie-algebra part, the consistency with Nambu--Lie commutation relation requires that this part should vanish. It is not clear how restrictive this condition is since not much is known on Nambu--Lie algebras in general, however in the case of $\A_4$ 3-algebra considered below this does not imply any additional constraints.

\subsubsection*{Irreducibility}
Constructing matrix representations of a Nambu--Lie algebra, one shall ask her/himself about the criteria of irreducibility of the representation. If we define an \emph{irreducible representation} as one for which \emph{the representation module has no invariant subspaces others than itself or zero element}, then if we restrict ourself to a subspace with definite value of $T_D$, the irreducibility of Nambu--Lie algebra representation follows directly from the irreducibility of the underlying Lie algebra representation.

It is worth noting that this definition also leads to a 3-algebra analog of \emph{Schur's Lemma}. Indeed, any operator which is Nambu commuting with any two arbitrary elements of the 3-algebra should commute in the sense of Lie algebra commutation with all generators of the underlying Lie algebra and, therefore, be proportional to the identity operator on each irreducible representation of the latter. That is not all, however. The Nambu bracket involving \emph{two} Lie algebra generators reads,
\begin{equation}
  0=[T_i,T_j,X]=[T_i,T_j]X=\ii X f_{ijk}T_k \Rightarrow X= 0,
\end{equation}
where we used the fact that $X$ is commuting with all generators $T_i$ following from the other Nambu bracket relations,
\begin{equation}
  0=[T_D,T_i,X]_{N}=[T_i,X]T_D+[X,T_D]T_i
  \Rightarrow [T_i,X]=0,
\end{equation}
for a irreducible underlying Lie algebra representation. Then the 3-algebra analogue of the Schur's lemma is somehow more restrictive:
\begin{lemma}
  There are no central elements in an irreducible representation of a Nambu Lie 3-algebra except the trivial one: $X=0$.
\end{lemma}

In particular this implies that there are no analogues of Casimir operators for 3-algebras.

\subsection{Algebra $\A_4$}
So far the only known nontrivial example of Euclidean finite-dimensional 3-algebra is the four dimensional algebra $\A_4$, which is generated by the Nambu--Lie commutation relations,
\begin{equation}\label{nambu4}
  [T_a,T_b,T_c]=\ii \epsilon_{abcd}T_d,
\end{equation}
where $\epsilon_{abcd}$ is the four-dimensional totally anti-symmetric tensor with $\epsilon_{1234}=+1$. Moreover, there is a strong evidence \cite{Takhtajan:1993vr,Gauntlett:2008,Papadopoulos:2008,Ho:2008a} that it is the only non-trivial Euclidean 3-algebra in finite dimensions.

As $\A_4$ is invariant with respect to SO(4) rotations,
all underlying Lie algebras are related by a SO(4) rotation and are isomorphic to su(2). Therefore, the choice of $T_4$ for the reduction of the Nambu--Lie bracket into a Lie algebra is a generic one. The underlying Lie algebra commutation relations are,
\begin{equation}\label{lie}
  [T_i,T_j]_{T_4}=-\ii \epsilon_{ijk}T_k, \quad i,j,k=1,2,3,
\end{equation}
where $\epsilon_{ijk}$ is the three-dimensional totally anti-symmetric tensor with $\epsilon_{123}=+1$.

On the other hand, the Nambu bracket commutation relations reduced to an eigen-space of $T_4$ transform to the following matrix commutator relations,
\begin{equation}\label{eq:t4su2}
  [T_i,T_j]=-\ii t_4^{-1}\epsilon_{ijk}T_k.
\end{equation}

Irreducible $(2j+1)$-dimensional representations of the Lie algebra generated by \eqref{eq:t4su2} are parameterized by the half-integer spin $j$. The Casimir operator is given by,
\begin{equation}
  T_i^2=t_4^{-2} \tau_i^2=t_4^{-2}j(j+1).
\end{equation}

Consideration of the leftover Nambu--Lie bracket $[T_1,T_2,T_3]$ gives the constraint on the value of $T_4$,
\begin{multline}\label{mul:t4}
  \ii t_4\equiv\ii T_4=[T_1,T_2,T_3]_{N}=T_1[T_2,T_3]+T_2[T_3,T_1]+T_3[T_1,T_2]\\
  =-\ii t_4^{-1} (T_1^2+T_2^2+T_3^2)=-\ii t_4^{-3} j(j+1),
\end{multline}
from which we have
\begin{equation}
  t_4^4=-j(j+1).
\end{equation}

As one can see it is not possible in this simple setup to represent the algebra $\A_4$ in terms of Hermitian matrices only.\footnote{
Let us note that this type of representations were first considered in \cite{Kawamura:2003cw} (see also \cite{Ho:2008a}), but it seems that the minus sign in front of $t_4^4$ was mishandled in both papers.}

\subsection{Nonconstant $T_4$ representation}

One may be concerned that the impossibility to find a unitary representation of $\A_4$ is related to the fact that we considered only the representations with a constant value of $T_4$. This implies that $T_4$ commutes with other three generators $T_i$ on each irreducible space. One may wonder whether giving up the commutativity of $T_4$ may improve the situation.

For non-commuting $T_4$ the Nambu commutator \eqref{nambu4} reduces to the following Lie bracket,
\begin{equation}\label{dcomm}
  \qp{A,B}_{T_4}\equiv T_4[A,B]+[A,B]T_4-AT_4B+BT_4A.
\end{equation}
One can consider this relation as a $T_4$ dependent deformation of usual commutation relations. Therefore, a $d_j$-dimensional representation in terms of bracket \eqref{dcomm} should be a deformation of an ordinary matrix representation of the underlying Lie algebra.

Consider the algebra $\A_4$. Here let us limit ourselves to the two-dimensional representation of the underlying algebra su(2). The deformed commutation relations in this case read,
\begin{equation}\label{mu-rep}
  \mu [\tau_i,\tau_j]+[\tau_i,\tau_j]\mu-\tau_i\mu\tau_j+\tau_j\mu\tau_i
  =\ii \epsilon_{ijk}\tau_k,
\end{equation}
where the generators $\tau_i$, $i=1,2,3$ are 2$\times$2-dimensional $\mu$-dependent matrices while $\mu\equiv  T_4$, for notational convenience.

To find a ``solution'' for the commutation relations \eqref{mu-rep} let us expand everything in terms two-dimensional Pauli matrices,
\begin{equation}
  \mu=\mu_0\I+\mu_\alpha\sigma_\alpha,\quad \tau_{i}=\tau_0\I+\tau_{i\alpha}\sigma_\alpha.
\end{equation}
Substituting this expansion into \eqref{mu-rep} yields,
\begin{equation}\label{sol1}
  2\mu_0\epsilon_{\alpha\beta\gamma}\tau_{i\alpha}\tau_{j\beta}\sigma_\gamma
  +6\mu_\sigma\epsilon_{\alpha\beta\sigma}\tau_{i\alpha}\tau_{j\beta}=
  \epsilon_{ijk}(\tau_{k\alpha}\sigma_\alpha+\tau_{k0}).
\end{equation}

The equation \eqref{sol1} is satisfied iff each of the following equalities hold,
\begin{equation}\label{sol2}
  2\mu_0\epsilon_{\alpha\beta\gamma}\tau_{i\alpha}\tau_{j\beta}
  =\ii \epsilon_{ijk}\tau_{k\gamma},\quad
  6 \epsilon_{\alpha\beta\gamma}\tau_{i\alpha}\tau_{j\beta}\mu_{\gamma}
  =\epsilon_{ijk}\tau_{k0}.
\end{equation}

The equalities \eqref{sol2} can be equivalently rewritten as,
\begin{equation}\label{equiv2su2}
  [\hat{\tau}_i,\hat{\tau}_j]=\ii \mu_0^{-1}\epsilon_{ijk}\hat{\tau}_k,\quad
  \hat{\tau}_i=\tau_{i\alpha}\sigma_\alpha/2
\end{equation}
and
\begin{equation}
  \tau_{k0}=\ft32\mu_0^{-1}\tr\hat{\mu}\hat{\tau}_k=
  3\mu_0^{-1}\mu_{\alpha}\tau_{k\alpha}.
\end{equation}

To complete the solution to the problem, let us note that for arbitrary matrix $\mu$ the solution for $\hat{\tau}_i$ satisfying \eqref{equiv2su2} is given up to a unitary transformation in terms of Pauli matrices: $\hat{\tau}_i=\sigma_i/(2\mu_0)$ and $\tau_{k0}=\ft32\mu_k/\mu_0^2$.
Plugging these results into the leftover Nambu commutator we obtain the equality,
\begin{equation}
  \mu=-\ii [\tau_1,\tau_2,\tau_3]=-\mu_0^{-3}\mu_k\hat{\tau}_k
  -\ft34\mu_0^{-3},
\end{equation}
from which we get,
\begin{equation}
  \mu_0^4=-3/4,\qquad \mu_k=0,
\end{equation}
i.e. the representation reduces to one considered in the previous subsection.

Taking the above results we may conjecture that complex values of $T_4$ are unavoidable in general.

\section{Discussion}

In this work we give a construction for irreducible representations of a Nambu--Lie 3-algebra in terms of representations of underlying Lie algebras. It appears that irreducibility of the Lie algebra automatically leads to the irreducibility of the entire Nambu--Lie 3-algebra representation.

Applied to the case of four-dimensional 3-algebra $\A_4$ the construction results in representations, which necessarily require matrices with eigenvalues which are fourth root of $-1$. This means that such representations can not be built entirely of hermitian matrices.

On the other hand, if we consider a Lorenzian 3-algebra which is obtained by the uplift procedure from the Lie algebra the mechanism of sign flip in the metric seem to be very similar to the sign flip in our construction. In this case one may expect that Lorenzian 3-algebras may be represented in terms of Hermitian generators. If it is the case, it is very much in contrast to what we have in Lie algebra representations.

\subsection*{Acknowledgements}
I thank Hai Lin for correspondence and finding typos in the manuscript as well as Jose A. de Azcarraga for drawing my attention to Generalized Lie algebras as an alternative to Nambu--Lie ones considered here. It is my pleasure also to thank Jeong-Hyuck Park and Bum--Hoon Lee for discussions and encouragement.

This work was supported by Center for Quantum Spacetime of Sogang University with grant number R11 - 2005 - 021.


\bibliographystyle{hunsrt}
\bibliography{BaggerLambert}
\end{document}